\documentclass[letters,fleqn,usenatbib]{mnras}

\usepackage{mathptmx}

\usepackage[T1]{fontenc}
\usepackage{ae,aecompl}

\usepackage{graphicx}	
\usepackage{amsmath}	
\usepackage{amssymb}	
\usepackage{times}
\usepackage{color}
\usepackage{float}
\usepackage{rotating}
\usepackage{url}

\def\apj{ApJ}
\def\mnras{MNRAS}
\def\aap{A\&A}
\def\apjl{ApJ}
\def\pasj{PASJ}

\def\iaucirc{IAU Circ.}

\def\apss{Ap\&SS}
\def\apjs{ApJS}

\def\aj{AJ}

\def\lesssim{\mathrel{\hbox{\rlap{\hbox{\lower4pt\hbox{$\sim$}}}\hbox{$<$}}}}
\let\la=\lesssim
\def\gtrsim{\mathrel{\hbox{\rlap{\hbox{\lower4pt\hbox{$\sim$}}}\hbox{$>$}}}}
\let\ga=\gtrsim

\title[The donor of Aquila X-1 revealed]{The donor of Aquila X-1 revealed by high angular resolution near-infrared spectroscopy}
\author[D. Mata S\'anchez et al.]{D. Mata S\'anchez $^{1,2}$\thanks{E-mail: dmata@iac.es}, T. Mu\~noz-Darias$^{1,2}$, J. Casares$^{1,2,3}$, F. Jim\'enez-Ibarra$^{1,2}$\\
$^{1}$Instituto de Astrof\'isica de Canarias, 38205 La Laguna, Tenerife, Spain\\
$^{2}$Departamento de astrof\'isica, Univ. de La Laguna, E-38206 La Laguna, Tenerife, Spain\\
$^{3}$Department of Physics, Astrophysics, University of Oxford, Denys Wilkinson Building, Keble Road, Oxford OX1 3RH, UK\\}

\date{Accepted 2016 August 27. Received 2016 August 26; in original form 2016 July 12}

\pubyear{2016}
\begin{document}
\label{firstpage}
\pagerange{\pageref{firstpage}--\pageref{lastpage}}
\maketitle

\begin{abstract}
The low mass X-ray binary Aquila X-1 is one of the most active neutron star X-ray transients. Despite its relatively bright quiescent optical counterpart, the detection of its companion has been hampered by the presence of a nearby interloper star. Using the infrared integral field spectrograph SINFONI on the VLT-8.2m telescope, we unambiguously single out Aquila X-1 from the interloper. Phase-resolved near infrared spectroscopy reveals absorption features from a $\rm{K}4\pm 2$ companion star moving at a projected velocity of $K_2= 136\pm 4 \, \rm{km \, s^{-1}}$. We here present the first dynamical solution and associated fundamental parameters of Aquila X-1, imposing new constraints on the orbital inclination ($36 \, ^{\circ}  < i < 47 \, ^{\circ}$) and the distance ($d = 6\pm 2\, \rm{kpc}$) to this prototypical neutron star transient.
\end{abstract}


\section{Introduction}
\label{intro}

Neutron star X-ray transients (NSXRTs) are a sub-type of low mass X-ray binaries harbouring a low mass star ($\lesssim 1 \,M_\odot$) and a neutron star (NS). They spend most of their lives in a faint, quiescent state, but show occasional outbursts where their X-ray luminosity increases above $\sim 10$ per cent of the Eddington luminosity. Observing NSXRTs in outburst, while convenient for accretion studies (e.g. \citealt{Munoz-Darias2014}), implies that most of the system luminosity arises from non-stellar components, completely veiling the companion star spectral features and preventing a dynamical solution even in the infrared (e.g. \citealt{MataSanchez2015}). On the other hand, the study of NSXRTs in their fainter, quiescent state -- where the relative contribution of the companion star to the total flux is larger -- depends greatly on the distance to the source and the Galactic extinction.

Aquila X-1 (hereafter Aql X-1, discovered by \citealt{Kunte1973}) is a recurrent NSXRT which has exhibited both coherent millisecond X-ray pulsations at $\sim 1.8\, \rm{ms}$ \citep{Casella2008} and thermonuclear bursts (e.g. \citealt{Galloway2008}). Despite showing recurrent outbursts and having a relatively accessible quiescent optical magnitude (V=21.6, \citealt{Chevalier1999}), a radial velocity study of the donor star is still missing. This has been prevented by the presence of an interloper star placed less than 0.5 arcsec apart from Aql X-1  \citep{Chevalier1999}. The interloper is $\sim 2$ mag brighter than Aql X-1 in the V-band but of comparable brightness at NIR wavelengths. In this work, we exploit the better spatial resolution inherent to the near-infrared (nIR) observations as well as adaptive optics techniques to obtain phase resolved, Integral Field Spectroscopy (IFS), which allow us to clearly resolve Aql X-1 from the interloper star. 

\section{Observations}
\label{observations}

\begin{figure*}
\includegraphics[width=190mm]{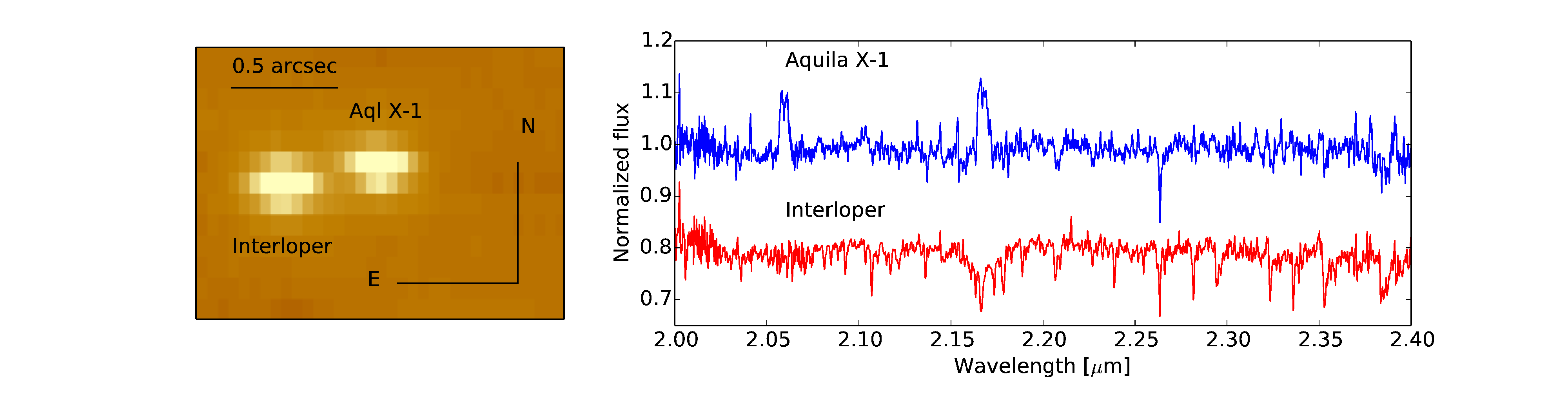}
\caption{Left panel: collapsed cube in the spectral dimension. Aql X-1 and the interloper are clearly resolved. Right panel: averaged, normalized spectrum of Aql X-1(blue, solid line) and the interloper star (red, solid line). A flux offset of 0.2 has been applied for display purposes.}
\label{figphot}
\end{figure*}

SINFONI is a nIR integral field spectrograph fed by an adaptive optics module currently installed at the Cassegrain focus of the 8.2m Very Large Telescope (VLT) at Cerro Paranal (Chile).
We obtained IFS in K-band (resolving power $R\sim 4000$) with an adaptive optics module working under natural guide star mode and a spatial pixel scale of $0.05''\times 0.1''$ (with a field-of-view of $3'' \times 3''$) 

The campaign includes 24 observations performed across several nights and sampling different orbital phases (always during the quiescent state) from 5 May 2010 to 29 September 2011. 
Each observation consists of 3-nodding $300\,\rm{s}$ exposures that were combined to subtract the sky contribution for a total exposure time of $900\,\rm{s}$ per observation. The reduction of the spectroscopic cubes was performed using  public SINFONI pipeline recipes.
The final products are 24 cubes, where Aql X-1 and the interloper star are clearly resolved (see Fig.\ref{figphot}) with a typical full-width-at-half-maximum (FWHM) of $\sim 0.18''$ (note that the K-band diffraction limit for an 8.2m telescope is $\sim 0.07''$). Early type standard stars were observed each night with the same instrumental configuration in order to carry out a telluric correction. Likewise, K0--M0 spectral type templates were obtained for radial velocity analysis and spectral classification.

We extracted the spectra using an aperture radius of $0.1''$, which provided independent, non-contaminated spectra of both Aql X-1 and the interloper. Spectra were analysed with the \textsc{molly} software. A telluric correction was applied using the atmospheric transmission profile derived from each standard star after masking their early type spectral features. Since our study requires precise measurement of orbital velocities, a wavelength correction due to flexure effects was performed using the sky spectra. The spectra were re-calibrated using OH emission lines \citep{Rousselot2000}, which yielded corrections under $<40\, \rm{km\, s^{-1}}$. The final, averaged spectra  are presented in Fig.\ref{figphot}.

\section{Results}
\label{results}

The averaged spectrum of Aql X-1 (Fig. \ref{figphot}) shows both emission lines from the accretion disc and absorption features from the donor star.  The individual spectra are noisier, but have enough quality [signal-to-noise ratio ($SNR$) $\sim 11$] to perform a radial velocity study.

\begin{figure*}
\includegraphics[width=190mm]{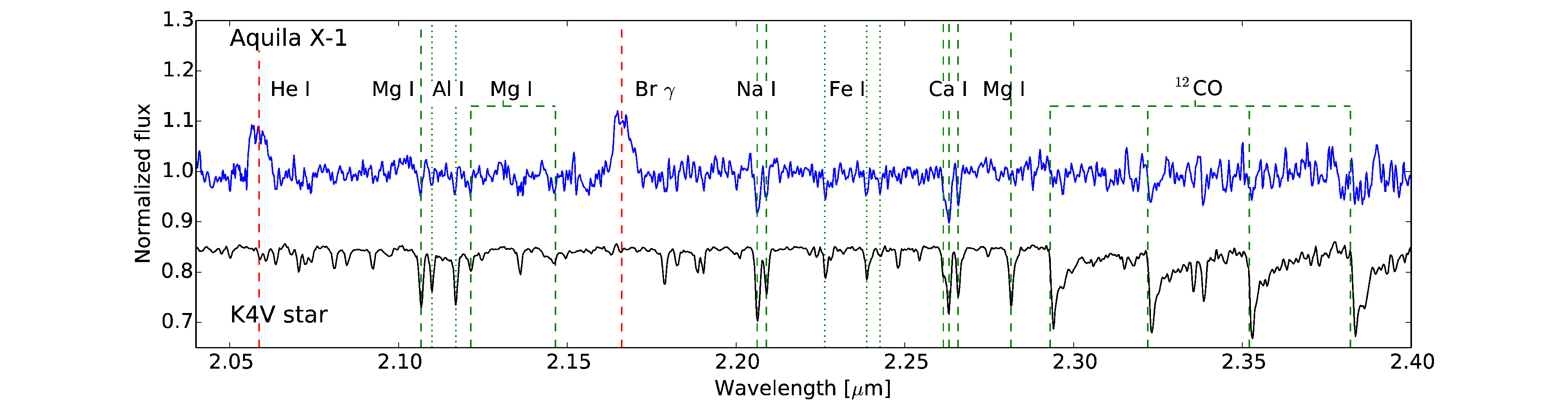}
\caption{Averaged spectrum of Aquila X-1 in the rest frame of the companion star (top). Accretion disc emission features (Br$\gamma$ and \ion{He}{I}) are marked with red, dashed vertical lines. The expected K-type donor star absorption features are shown as green, dashed and dotted vertical lines, with the \ion{Na}{I} doublet and the \ion{Ca}{II} triplet being the most prominent. A template $\rm{K4V}$ spectrum is depicted for comparison (bottom).
}
\label{figaverage}
\end{figure*}

\begin{figure}
\includegraphics[width=90mm]{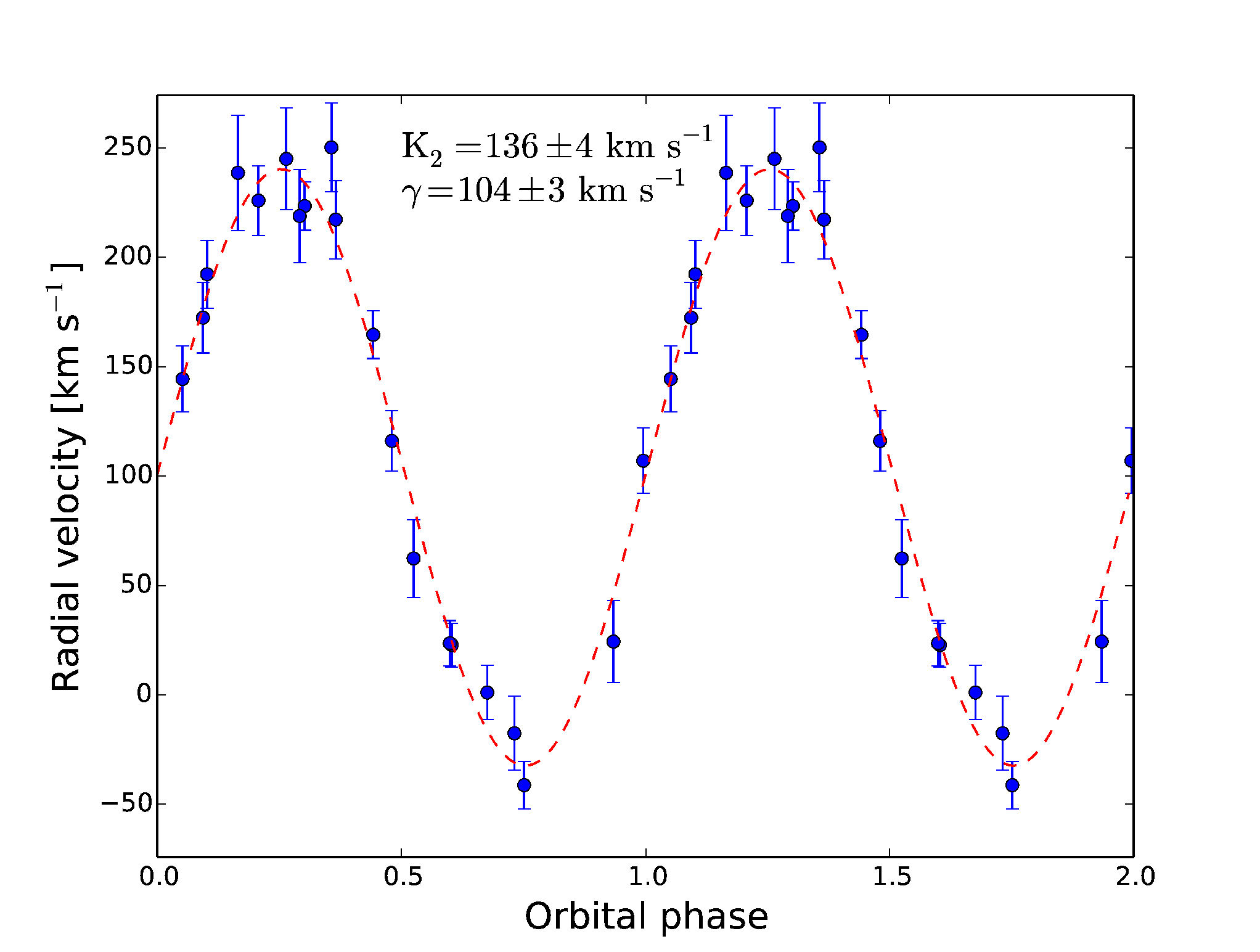}
\caption{Radial velocity curve obtained using a $\rm{K}4\rm{V}$ template. A sinusoidal fit is depicted as a dashed red line. Four spectra whose radial velocities deviate more than $1.5\sigma$ from the fit due to their lower SNR (poor seeing) were discarded in order to produce a cleaner radial velocity curve. Two orbital phases are shown for clarity.}
\label{figK2rv}
\end{figure}

\subsection{Radial velocity curve of the donor star}

Our SINFONI spectral templates have a SNR too low to be used in a radial velocity study. Instead, we compared our object spectra with F to M main-sequence stars templates from the \textit{IRTF Spectral Library} (\citealt*{Rayner2009}, \citealt*{Cushing2005}) after degrading our data to their lower resolution ($R\sim 2000$). We use the \textsc{molly} \textit{xcor} routine to perform the cross-correlation, obtaining the radial velocity shift of each individual spectrum. To carry out this task we considered a spectral region free of both emission lines and residuals from the telluric correction, where prominent absorption features such as \ion{Na}{I} doublet ($2.206-2.208 \, \rm{\mu m}$) and \ion{Ca}{I} triplet ($2.261-2.266\, \rm{\mu m}$) are present (Fig. \ref{figaverage}). For each spectrum we computed the binary phase using the most accurate orbital elements available in the literature i.e. $P_{\rm{orb}}= 0.789498 \pm 0.000010\, \rm{d^{-1}} $ \citep{Garcia1999} and $T_0=2450282.220 \pm 0.003\, \rm{d}$ \citep{Chevalier1998}. We note that the phase uncertainty, projected onto our last data point, amounts to $\pm 0.10$. The resultant radial velocity curve (Fig. \ref{figK2rv}) reveals a clear orbital modulation caused by the motion of the donor star (see Sec. \ref{discussion}).  A sinusoidal fit to the data of the form:
\begin{equation}
\label{eq1}
V=\gamma +K_2\, \sin{(2\pi\, (\phi-\phi_0))}$$
\end{equation}
where $V$ is the observed velocity, $\gamma$ is the systemic velocity, $\phi_0$ the phase offset and $K_2$ the donor star orbital velocity, yields: $K_2= 136\pm 4 \, \rm{km \, s^{-1}}$; $\gamma = 104\pm 3 \, \rm{km \, s^{-1}}$ and $\phi_0= 0.130\pm 0.005$ ($\chi^2_{\rm{red}}=0.86$). This is obtained from the cross-correlation with the $\rm{K4V}$ template, but we note that G to M templates return values consistent within $\rm{1\sigma}$. Our data allows us to define a new $T_0$ from the epoch with the maximum number of observations ($T_0=2455810.387\pm 0.005 \rm{d}$). Using  this new value we  refine the orbital period to $P_{\rm{orb}}= 0.7895126\pm 0.0000010 \, \rm{d^{-1}}$.

\subsection{Spectral classification and veiling}

Fig. \ref{figaverage} shows that both the donor star and the accretion flow contribute to the nIR emission of Aql X-1 . Therefore, the actual spectrum of the companion is veiled by a factor $X$, defined as the fractional contribution of non-stellar sources to the total flux (e.g. the accretion disc contribution). In order to constrain $X$ and the spectral type of the companion we produced, for each template, a grid of spectra sampling the range $0<X<1$, and compared them with the averaged spectrum of Aql X-1 in the donor star rest frame using a $\chi^2$ minimization routine. The minimum of the $\chi^2$ distribution constrains the donor star spectral type to be $\rm{K}4\pm2$ and the K-band veiling to $X=0.36\pm 0.10$. The same analysis reveals a $\rm{G9\pm 2V}$ spectral type for the interloper.

\subsection{Diagnostic diagram: The radial velocity curve of the NS}

In order to constrain the radial velocity of the neutron star we analysed the evolution of the most prominent emission line present in the spectra, namely  Br$\rm{\gamma}$ ($2.166\,  \mu m$). We constructed a diagnostic diagram \citep{Shafter1986} where we fit the emission line in each spectrum to a double Gaussian model with a fixed separation ($a$). Large $a$ values trace the movement of the wings of the line, which are expected to be formed in the inner parts of the accretion disc. For each $a$ value we apply a sinusoidal fit (similar to eq. \ref{eq1}) to the obtained velocities. By averaging the results for separations in the range $a=600-800 \, \rm{km\, s^{-1}}$, where we found that the fit was acceptable (including visual inspection) and stable, we infer: $K_1=56\pm 11\, \rm{km\, s^{-1}}; \, \gamma= 132\pm 10\, \rm{km\, s^{-1}}; \phi_0=0.63\pm0.03$. We carried out the same analysis using the fainter \ion{He}{I}  ($2.059\, \mu m$) line and obtained consistent results.

The systemic velocity derived from this technique is consistent within $2.2\rm{\sigma}$ with the $\gamma$ derived from the donor star radial velocity curve. Likewise, the zero phase $\phi_0 \sim 0.5$ suggests that the wings of the lines are formed in the surrounding disc around the NS, tracing its movement. Therefore, we propose $K_1= 56\pm 11\, \rm{km\, s^{-1}}$ as a good estimate of the radial velocity of the neutron star. We note that both $K_1$ and $\gamma$ derived from the diagnostic diagram are consistent with previous determinations using optical emission lines by \citeauthor{Garcia1999} (\citeyear{Garcia1999}, $K_1=70\pm 20\, \rm{km\, s^{-1}}$,$\gamma=150\pm 20\, \rm{km\, s^{-1}}$, H$\rm{\alpha}$) and \citeauthor{Cornelisse2007}  (\citeyear{Cornelisse2007}, $K_1=68\pm 5\, \rm{km\, s^{-1}}$, $\gamma=130\pm 20\, \rm{km\, s^{-1}}$, H$\rm{\beta}$).

\subsection{Astrometry of Aql X-1 and light curves}

\begin{figure}
\includegraphics[width=90mm]{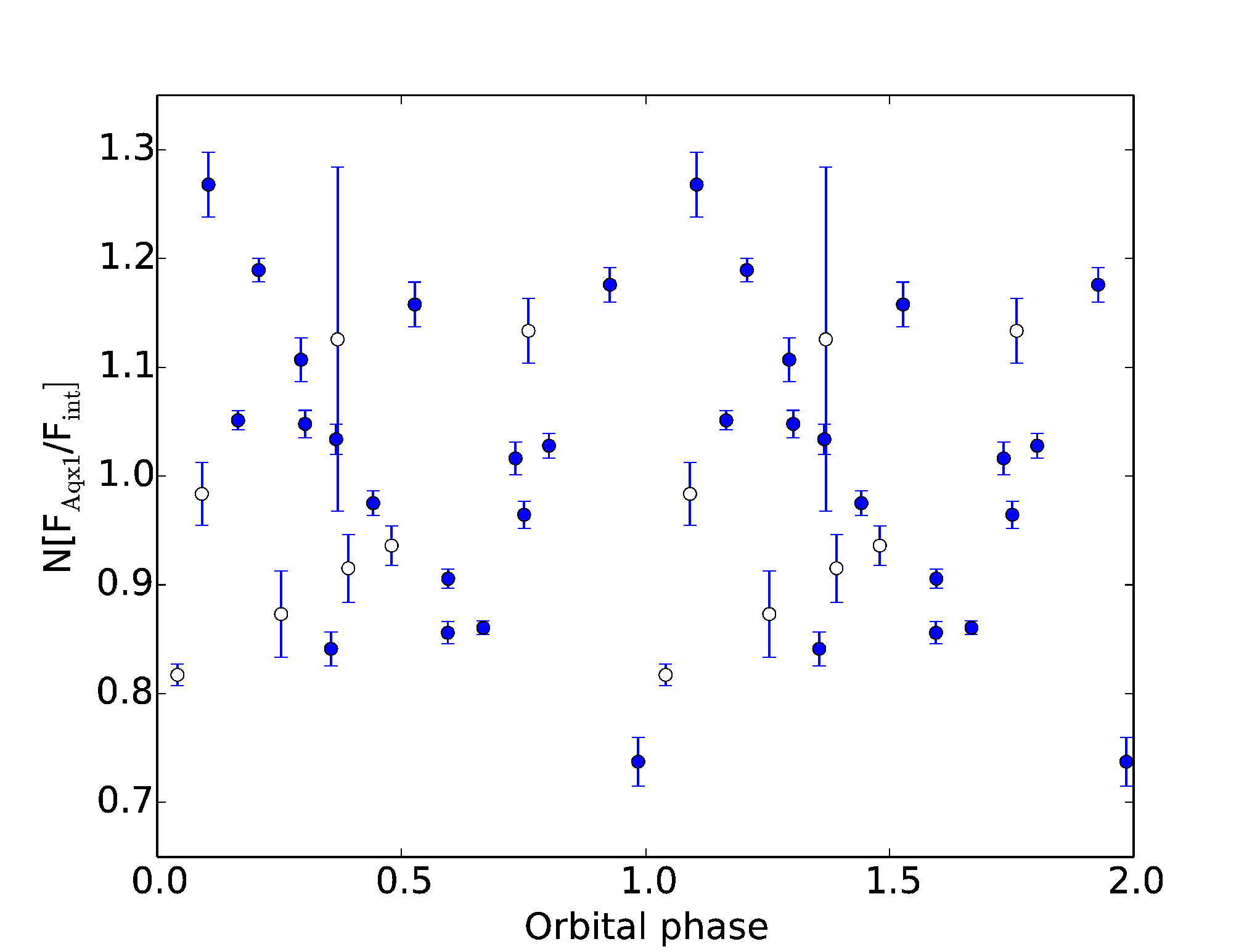}
\caption{K-band flux variability of Aql X-1 (relative to the interloper) as a function of the orbital phase. Filled dots refer to observations with seeing better than $<0.22''$, while empty dots depict those with poorer seeing (up to $0.40''$). Two orbital cycles are shown for clarity.}
\label{figlight}
\end{figure}

The high angular resolution achieved by SINFONI allowed us to single out Aql X-1 from a nearby field star. Aiming at refining the object coordinates, we collapsed each individual cube along the spectral dimension and fitted a bidimensional Gaussian to the resultant "photometric" images. We derived a centroid position relative to the interloper of $\rm{\Delta \alpha = -0.0295\pm 0.0006\, s}$; $\rm{\Delta \delta= 0.083\pm 0.011\, ''}$, which is consistent within $1.5\sigma$ with that obtained from ultraviolet observations \citep{Hynes2012}, yielding a separation of $\rm{ 0.450\pm 0.014\, arcsec}$. The collapsed images can also be used to perform differential photometry against the interloper star, which is expected to remain constant during the observations. The obtained light curve from aperture photometry (see Fig. \ref{figlight}) shows variability at $\sim 20\%$ ($\sim 0.3\, \rm{mag}$) level ($\sigma = 0.13$), but there is no clear orbital modulation (see  Sec. \ref{discussion}).

\subsection{Rotational broadening}

In an attempt to measure the rotational broadening of the companion star in Aql X-1 we compared the breadth of the absorption features -- using the averaged spectrum in the companion star rest frame -- with template stars observed with the same instrumental configuration. These cover spectral types K0IV, K3V, K7V and M0V and were co-added in order to produce a single "K-type" template with sufficient SNR for this analysis. We then artificially broadened this grand-sum template spectrum using a linear limb-darkening coefficient of 0.33 \citep{AlNaimiy1978}, and compared it with the Aql X-1 spectrum using \textit{optsub} \textsc{molly} routine to determine the best fit. Unfortunately, we were not able to resolve the minimum of the $\chi^2$ distribution. This is not surprising given that our spectral resolution is $\sim 75\, \rm{ km\, s^{-1}}$ and we expect (using our $K_2$ and $K_1$ values; see \citealt{Wade1988}) a rotational broadening of $\sim 60\,  \rm{ km\, s^{-1}}$ . 

\section{Discussion}
\label{discussion}

We have determined the orbital velocities of the system components in Aql X-1 as well as the donor star spectral type. We can use these results to constrain other fundamental parameters of the system such as the inclination, the NS mass and the distance.

\subsection{The distance to Aql X-1}

K-band photometry was carried out using the data cubes after collapsing in the spectral dimension. These were compared with standard stars observed each night. The mean values for the interloper star and Aql X-1 are $K_{\rm{Int}}=16.3\pm 0.3$ and $K_{\rm{Aql}}=16.7\pm 0.3$, respectively, where the uncertainty refers to one standard deviation. The observed hydrogen equivalent column density $N_{\rm{H}}=( 4.4\pm0.1) \times 10^{21}\, \rm{cm^{-2}}$ \citep{Campana2014} and the relation $N_{\rm{H}}=( 2.0\pm 0.5) \times 10^{21}\, \rm{cm^{-2}} A_{\rm{V}}$ \citep{Watson2011} yields $A_{\rm{V}}=2.2\pm 0.6$. We apply the extinction relation provided by \citet{Cardelli1989} to derive the extinction in the K-band: $A_{\rm{K}}=0.114\, A_{\rm{V}}=0.25\pm 0.06$. If we consider the veiling factor from Sec. \ref{observations}, the apparent magnitude of the donor star of Aql X-1 is $K_{\rm{donor}}=16.9\pm 0.4$.

\citet{Cox2000} provides absolute visual magnitudes and stellar parameters of dwarf and giant stars. Conversion to the K-band can be achieved through the bolometric correction tables from \citet{Masana2006}. On the other hand, a $\rm{K}4\pm2$ dwarf star is not large enough to fill the Roche Lobe of a 19h orbital period system, and therefore no accretion would be expected. The frequent outbursts observed in Aql X-1 require the donor star to be an evolved star which has expanded to the Roche Lobe radius $R_{\rm{L}}=1.5\pm 0.1\, R_\odot$ \citep{Eggleton1983}. Taking this into account, we predict an absolute magnitude for the donor of Aql X-1 of $M_{\rm{K}}=3.0 \pm 0.8$, which results in a distance of $d  = 6\pm 2\, \rm{kpc}$. Our result is consistent with that derived from thermonuclear X-ray burst analysis ($d<6\,\rm{kpc}$, \citealt{Galloway2008}). If we consider the interloper star to be a $\rm{G9\pm 2 V}$ and the extinction to be lower than that of Aql X-1, it places this object at $d_{\rm{Int}} = 2.0 - 4.1\, \rm{kpc}$, in agreement with previous determinations \citep{Chevalier1999}.

The radial velocity curve of the Galaxy reported in \citet{Clemens1985} and refined by \citet{Nakanishi2003} implies that the expected radial velocity of an object at the position of Aql X-1 would be $68-92\,\rm{km\, s^{-1}}$ ($42-65\,\rm{km\, s^{-1}}$ for the interloper star). From our spectra we found a systemic velocity for the interloper of $43 \pm 7 \,\rm{km\, s^{-1}}$, which is fully consistent with the expected value. In the case of Aql X-1 we measure $\gamma =104\pm 3\, \rm{km\, s^{-1}}$. This implies a natal kick in the radial direction of $12-36 \,\rm{km\, s^{-1}}$, in agreement with findings in others neutron star binaries (e.g. Cen X-4, \citealt{GonzalezHernandez2005}).

\subsection{Masses and inclination of Aql X-1}

Our radial velocity curve, presented in Sec. \ref{results}, provides the first dynamical detection of the donor star in Aql X-1, revealing a projected velocity of $K_2=136\pm 4\, \rm{km\, s^{-1}}$. Previous attempts to measure the donor star orbital velocity in the optical regime were hampered by the presence of the interloper star. The so-called Bowen technique, which studies narrow emission lines (particularly the Bowen blend) arising from the inner, irradiated face of the companion star during the outburst, proposed $K_2\sim 250\, \rm{km\, s^{-1}}$ \citep{Cornelisse2007}. Neither the obtained velocity amplitude (see e.g. \citealt{Munoz-Darias2005}) nor the orbital phases are compatible with the solution presented here. Although the Bowen technique has proven to be valid in several systems (e.g. \citealt{Steeghs2002}), a contaminating contribution to the Bowen blend by structures arising in other parts of the binary different from the irradiated face of the companion (e.g. hot-spot) has been observed in some cases (e.g. Ser X-1, \citealt{Hynes2004}). This issue (as noted by the authors in \citealt{Cornelisse2007}) together with a limited orbital coverage probably concealed the donor star emission features in their spectra. A full optical spectroscopic study of Aquila X-1 in outburst will be presented in a forthcoming work (Jim\'enez-Ibarra et al. in prep.).

The results derived in Sec. \ref{results} ($K_2$, $P_{\rm{orb}}$ and $K_1$) produce a mass ratio $q=\frac{M_{\rm{donor}}}{M_{\rm{NS}}}=\frac{K_1}{K_2}=0.41\pm 0.08$, as well as a mass function:
{\begin{equation}
f(M_{\rm{NS}})=\frac{M_{\rm{NS}}\, \sin{i}^3}{(1+q)^2}=\frac{P_{\rm{orb}}\, K_2^3}{2\pi\,G}=0.21\pm 0.02 ,
\label{eqmassfunct}
\end{equation}
which does not significantly restrict the mass of the NS due to both the uncertainty in $q$ and the unknown value of the orbital inclination ($i$). The detection of both thermonuclear burst and pulsed emission indicates a NS acretor in Aql X-1, and thus its mass should be higher than $M_{\rm{NS}}>1.2\, M_\odot$ \citep{Kiziltan2013}. Similarly, the companion star has been classified as an evolved $\rm{K}4\pm 2$ star; assuming that binary evolution does not significantly increase its mass (see \citealt*{Kolb2001}) then $M_{\rm{donor}}< 0.76\, M_\odot$ \citep{Cox2000}, which all together yields an upper limit to the mass ratio of $q<0.63$. Using our $K_2$ and $P_{\rm{orb}}$, and considering a conservative constraint of $M_{\rm{NS}}< 3\, M_\odot$, the inclination parameter is restricted to $23 \, ^{\circ} < i < 53 \, ^{\circ}$ (see Fig. \ref{figmass}). By including the proposed $K_1$ and the upper limit imposed by the most massive NS found so far ($M_{\rm{NS}} \sim 2\, M_\odot$, PSR J0348+0432; \citealt{Antoniadis2013}), we obtain $34 \, ^{\circ} < i < 47 \, ^{\circ}$. On the other hand, photometric observations \citep{Welsh2000} constrain the orbital inclination to $i > 36\rm{^{\circ}}$, which is fully consistent with our results. A canonical NS ($M_{\rm{NS}}=1.4\, M_ {\odot}$) would require $i=42 \pm 3\, ^{\circ}$. \citet{Galloway2016} have very recently reported on the presence of intermittent X-ray dips on Aql X-1, which seem to be only present in $\sim$ 10\% of the orbital cycles. Our less constraining (i.e. more solid) upper limit ($i<53^{\circ}$ for $M_{\rm{NS}}>1.2\, M_\odot$) rules out a high orbital inclination as the origin of these events.

\begin{figure}
\includegraphics[width=90mm]{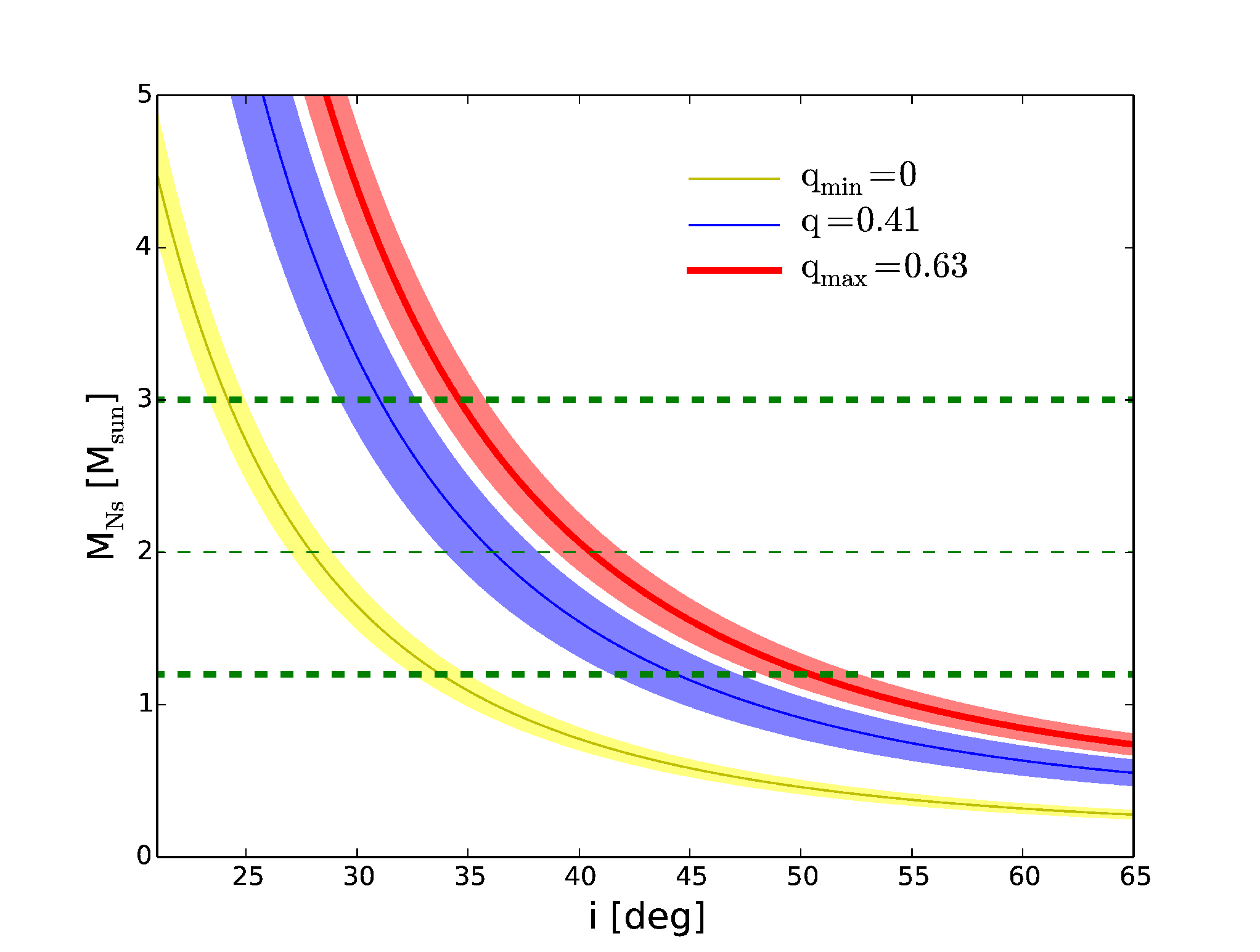}
\caption{Neutron star mass versus orbital inclination. Horizontal, thick dashed lines correspond to the limits imposed to the NS mass (1.2 and 3 $M_\odot$), whereas the thin dashed line marks the most massive NS found so far ($ 2\, M_\odot$; \citealt{Antoniadis2013}). The three solid lines represent our extreme (yellow and red lines) and preferred q values (blue line). The uncertainties (derived from $K_2$ and $P_{\rm{orb}}$) are plotted as shaded regions.}
\label{figmass}
\end{figure}

\citet{Casares2015} found a linear correlation between $K_2$ and the FWHM of the H$_\alpha$ line. The more general solution presented in the aforementioned paper,
\begin{equation}
\label{eqK2FWHM}
\frac{K_2}{\rm{FWHM}}=\frac{\sqrt{\alpha f(q)}}{2}; \qquad f(q)=\frac{0.49(1+q)^{-1}}{0.6+q^{2/3}\ln (1+q^{-1/3})}
\end{equation}
assumes that the FWHM of the emission line is determined by the gas velocity at a characteristic disc radius $R_{\rm W}=\alpha R_{\rm{L}1}$, where $R_{\rm{L}1}$ is the distance from the compact object  to the Lagrangian point $\rm{L}_1$. This empirical correlation implies that the H$_\alpha$ line is typically formed (in quiescence) at about $0.42R_{\rm{L}1}$. Following the same procedure but using Br$_{\gamma}$ instead of H$_\alpha$ together with our $q$ and  $K_2$ values suggest that this line is formed at more external radii ($0.54\pm 0.11 \times R_{\rm{L}1}$), which is an expected feature supporting our results.

Finally, we note that the quiescent nIR emission of  Aql X-1 shows significant long-term variability at 20\% level, but no apparent orbital modulation (Fig. \ref{figlight}). The latter could be due to both the relatively low inclination that we have derived and the presence of accretion related variability during the quiescent state (our observations were spanned over different epochs across more than a year). Spurious variability owing to a variable contribution from the interloper star is not expected since it only typically accounts for less than 5 per cent of the object flux (seeing $\sim 0.2’’$; see Fig. \ref{figlight}). On the other hand, X-ray variability during the quiescence has been observed in Aql X-1 \citep{Cackett2011}, and optical variability is a common property of both quiescent black hole and NS transients (e.g. \citealt{Zurita2003}).

\section{Conclusions}
\label{conclusion}

We used near infrared integral field spectroscopy to single out Aquila X-1 from a nearby interloper star. The spectra reveal, for the first time, absorption features corresponding to a $\rm{K}4\pm2$ donor, veiled by $\sim 36\%$ in the K-band, and moving at a projected orbital velocity of $K_2=136\pm 4\, \rm{km\, s^{-1}}$. We further refine the ephemerides of the system to $T_0=2455810.387\pm 0.005 \rm{d}$ and $P_{\rm{orb}}= 0.7895126\pm 0.0000010 \, \rm{d^{-1}}$, and constrain the orbital inclination to $36 \, ^{\circ} < i < 47 \, ^{\circ}$. Using the de-reddened K-band magnitude and the constraints on the spectral type we infer a distance to the source of $d = 6\pm 2\, \rm{kpc}$.

\section*{Acknowledgements}
\label{acknowledgements}
DMS acknowledges Fundaci\'on La Caixa for the financial support received in the form of a PhD contract. We also acknowledge support by the Spanish Ministerio de Econom\'ia y competitividad (MINECO) under grant AYA2013-42627. JC is supported from the Leverhulme Trust through the Visiting Professorship Grant VP2-2015-046. Based on observations made with ESO telescopes at the La Silla Paranal Observatory under program ID 085.D-0271. MOLLY software developed by T. R. Marsh is gratefully acknowledged. 

\bibliographystyle{mnras} 
\bibliography{MiBiblio.bib}


\bsp	
\label{lastpage}
\end{document}